\documentclass[english,aps,preprint,nofootinbib]{revtex4}
\usepackage[]{fontenc}
\usepackage{graphicx}
\usepackage{amssymb}

\makeatletter

\usepackage{babel}
\makeatother
\begin{document}

\preprint{BROWN-HET-1395, NSF-KITP-04-25}

\title{Holography and Eternal Inflation}

\author{David A. Lowe}

\email{lowe@brown.edu}

\affiliation{Physics Department, Brown University, Providence, RI 02912, USA}

\author{Donald Marolf}

\email{marolf@physics.ucsb.edu}

\affiliation{Physics Department, University of California, Santa Barbara, CA 93106,
USA}

\begin{abstract}
We show that eternal inflation is compatible with holography. In particular, we emphasize that
if a region is asymptotically de Sitter in the future, holographic arguments by themselves place
no bound on the number of past $e$-foldings. We also comment
briefly on holographic restrictions on the production of baby universes.
\end{abstract}
\maketitle

\section{introduction}

New inflation models typically suffer from severe fine tuning problems
associated with the choice of initial state. In the eternal inflation
scenario, these problems are avoided because inflating bubbles persist
along any time-slice, and these inflating regions self-reproduce leading
to a fractal multiverse spacetime structure \cite{Linde:1994xx}.
From the point of view of a local observer the details of this multiverse
structure are irrelevant, except to set up the initial conditions
for the observer's own inflating bubble. 

In this paper, we examine constraints on the eternal inflation scenario
arising from holographic entropy bounds. Historically, the idea of
holographic bounds \cite{tHooft,LS} and their cousins \cite{Bek73,Bek2000a,Bek2000b,BV}
emerged from the study of black hole entropy (but see \cite{MS1,MS2,ObsEnt})
and some researchers find motivation for such bounds in certain results
from string theory. Such bounds are equivalent to the assumption that
black holes are maximally entropic objects of a given size; they state
that the entropy residing inside the relevant region is bounded by
its surface area in Planck units \cite{tHooft,LS,Easther:1999gk}.

On the other hand, for at least one proposed form \cite{Bousso} of
the holographic bound, it was argued in \cite{FMW} that when i) the
number of fields is small, ii) the matter $T_{\mu\nu}$ is not
too anisotropic locally, and iii) temperatures are below the Planck scale,
the bound follows as a consequence of the Einstein equations. Similarly,
one generally does not expect holographic bounds to impose additional
restrictions on thermodynamics at temperatures below the Planck scale.
However, Banks and Fischler argued that holography (of a somewhat different form than that used in \cite{Bousso}), together with certain additional assumptions, requires any late-time observer entering a region dominated by a small
value of the cosmological constant to observe a bounded
number of $e$-foldings \cite{Banks:2003pt}. See \cite{Wang:2003qr,Cai:2003zs}
for subsequent related works. 

Here we wish to emphasize one of the additional assumptions of \cite{Banks:2003pt}.  In particular, 
\cite{Banks:2003pt} considers a scenario where the universe is inflating at early times, passes through a matter-dominated regime, and then becomes asymptotically-de Sitter in the future.  The assumption of interest is that 
{\it the total entropy of the universe in the early-time inflationary region can be computed by local field theory methods 
even when no observer can directly measure all of this entropy.}  In particular, we will
see in section \ref{bound} that most of this entropy lies outside the past light cone of any observer. 

We are motivated to question this assumption by the observation that a similar assumption in the late-time de Sitter region would already violate any holographic bound on the entropy of the system.
This is just the observation that de Sitter space expands to infinite size in the far future, so that any field theory with any finite cutoff contains an infinite number of degrees of freedom.  This observation is not new, and is well known to proponents of holography who propose that nevertheless de Sitter space is associated only with a finite number of states.  The usual resolution (see e.g. \cite{Mobs}) is to note that this calculation does not contradict the experience of any observer in the spacetime, as such observers have access to only a small part of the entropy -- small enough, it turns out, to satisfy a holographic bound.  One then supposes that the true entropy of the system is comparable to the maximum entropy measurable by any given observer, and that field theory breaks down on scales large enough that it would predict violations of the holographic entropy bounds (see e.g., \cite{CKN}).

Our goal here is to show that applying similar reasoning to the system considered in  \cite{Banks:2003pt} yields a similar conclusion.  That is, in contrast to \cite{Banks:2003pt} 
we assume that holographic bounds restrict only the field theory entropy in any past light cone, as field theory may generally acquire holographic corrections on larger scales. In this context we show that holography imposes no restrictions on inflation.  In particular, the number of $e$-foldings can be arbitrarily large.
To distinguish our assumption from that 
of \cite{Banks:2003pt} we refer to it as ``light-cone holography'' below\footnote{Bousso's covariant entropy bound \cite{Bousso} a special case of light-cone holography, but we allow much more general settings here.   We emphasize that some form of light-cone holography is essential for {\it any} consistent holographic description of de Sitter space.}.
We also comment briefly on claims
\cite{BanksFalseVac} of holographic restrictions on baby Universe
formation.


\section{Bound on $e$-foldings from entropy in a light cone?}
\label{bound}

\begin{figure}
\includegraphics[%
  scale=0.75]{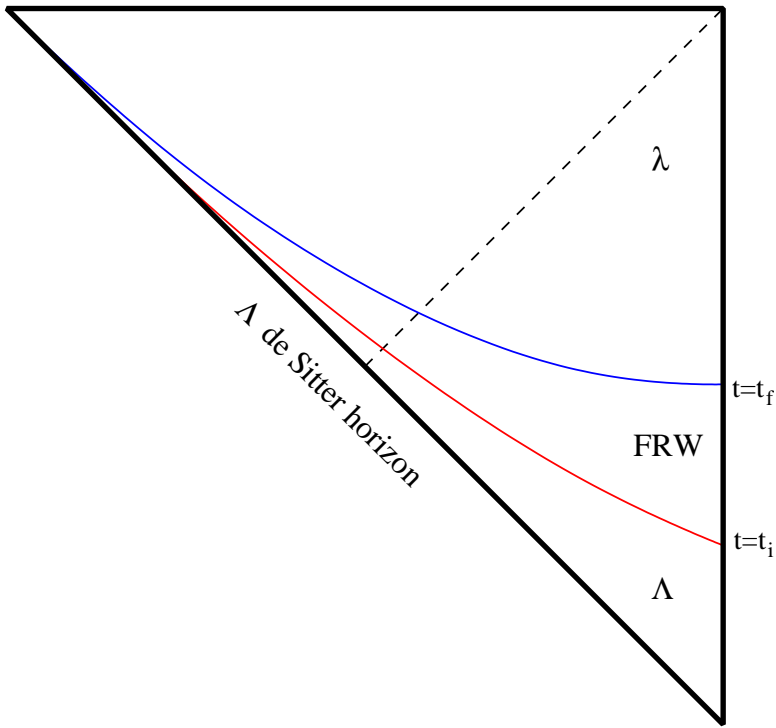}

\caption{\label{cap:Spacetime-diagram} Spacetime diagram. The dashed line
is the past light cone of a late-time observer.}
\end{figure}

\begin{figure}
\includegraphics[%
  scale=0.75]{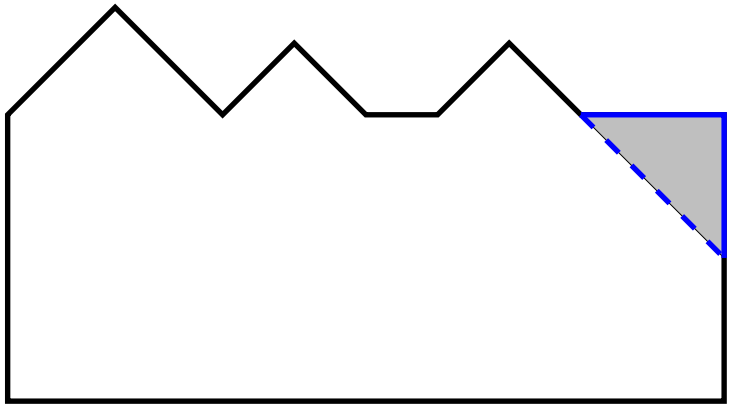}

\caption{\label{cap:A-representation-of}A representation of the spacetime
of eternal inflation. The shaded triangle corresponds to the region
shown in figure \ref{cap:Spacetime-diagram}.}
\end{figure}

In figure \ref{cap:Spacetime-diagram} below, we show a (rough) conformal
diagram of the class of spacetimes considered in \cite{Banks:2003pt}. This
spacetime region might appear as some portion of an eternally inflating
spacetime, which we show in over-simplified form in figure \ref{cap:A-representation-of}. An inflating region with vacuum energy density $\Lambda$,
is patched onto a Friedmann-Robertson-Walker phase dominated by some
form of matter satisfying $p=\kappa\rho$, which in turn asymptotes
to a de Sitter region with small cosmological constant $\lambda\ll\Lambda$.
We assume homogeneity, isotropy, and spatial flatness. The latter
is at least a good approximation as we are most interested in the
case where the $\Lambda$-region undergoes a large number of $e$-foldings.
Following \cite{Banks:2003pt}, we consider only the region shown
in figure \ref{cap:Spacetime-diagram} and do not concern ourselves
with holographic restrictions on what occurs to the past of the $\Lambda$-de
Sitter horizon.

The basic idea of our analysis is already clear from this diagram.
Consider some observer in the far future. Her past light cone to the
future of $t=t_{I}$ is determined by propagating null rays (dashed
line) backwards through the FRW and $\lambda$ regions. Clearly then,
the spacetime region visible to her but later than time $t_{i}$ is
independent of the number $N$ of $e$-foldings that takes place in
the $\Lambda$ region. Thus light-cone holographic bounds in the FRW region cannot
impose any restrictions on $N$. 

\begin{figure}
\includegraphics{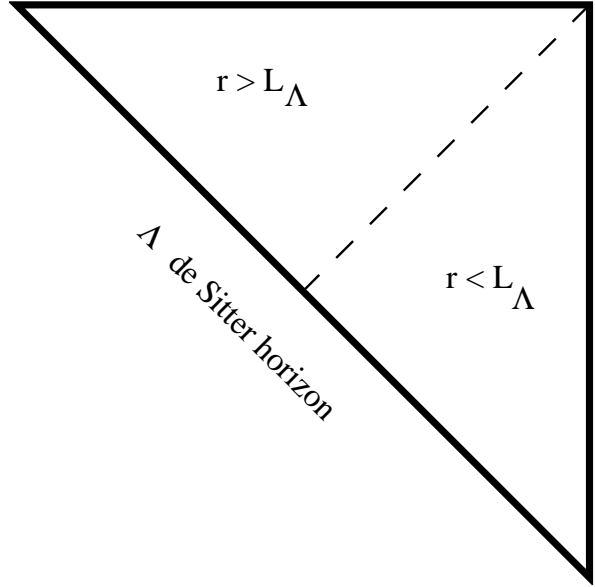}

\caption{\label{cap:The-inflating-patch}The inflating patch of $\Lambda$-de
Sitter space.}
\end{figure}

Now, if we continue to trace the light-cone backwards through the
$\Lambda$ region, we will find one of two things to be true. The
first possibility is that the light cone at $t=t_{i}$ is larger than
the de Sitter horizon in the $\Lambda$ region. For this discussion
the reader may wish to consult figure \ref{cap:The-inflating-patch}, which depicts the inflating patch of pure $\Lambda$-de Sitter
space with its horizons. We have also indicated in each region the
size of the suppressed spheres relative to $L_{\Lambda}$, the $\Lambda$-horizon
scale. In this first case, standard results from de Sitter space tell
us that the light cone \emph{contracts} (when traced backwards) until
we reach the de Sitter horizon. Thus, the largest piece of $\Lambda$-de
Sitter spacetime is seen at time $t=t_{i}$ and, since holographic
bounds are most stringent for large regions, light-cone holographic considerations
at earlier times can yield only weaker restrictions. In particular,
since a bound applied at $t=t_{i}$ cannot restrict $N$, it is impossible
for a bound at earlier times to do so. More generally, it is clear
that $t=t_{i}$ places the most stringent bound on the entropy. 

If on the other hand the light cone is smaller than the $\Lambda$-de
Sitter horizon at $t=t_{i}$, it will be expanding and will continue
to expand when traced further backward. However, it will remain smaller
than the horizon size until it crosses the $\Lambda$de Sitter horizon.
But it is well known that observing a horizon-scale region of de Sitter
space does not contradict any holographic bounds. Thus, once again
nothing new is learned from the region $t<t_{i}$. 

Thus it is clear that light-cone holographic considerations can place no bound
on $N$. Nonetheless, one might still ask whether they place bounds
on other quantities relevant to the scenario above. We now turn to
this question and investigate in detail the past light-cone of a late-time
observer.

\subsection{The late-time past light cone}

There is, however, the issue of just which sort of holographic bound
we wish to consider. One popular formulation is Bousso's covariant
entropy bound \cite{Bousso}. However, this counts only entropy that
flows through \emph{contracting} light-sheets and some holographers may
desire a tighter bound%
\footnote{Though it is not clear to what extent one could hold in general.%
}. Thus, for the rest of this
section we will simply assume that holographic considerations restrict
the entire entropy visible to any observer on any homogeneous spacelike
slice $\Sigma$ to be less than the area $A$ of the intersection
of $\Sigma$ with the observers past light cone.

Let us also pause to further orient ourselves to the problem at hand.
We have already seen that our light-cone holographic bound is satisfied in the
$\Lambda$-region if it is satisfied at $t=t_{i}$. It is also satisfied
in the $\lambda$-region by the usual arguments for de Sitter space.
To address the rest of the FRW region, consider the light cone at
$t=t_{i}$. Because the universe as a whole is homogeneous and expanding,
if the light cone were at any point expanding toward the future it
would continue to do so for all time and would not converge at the
location of our late-time observer. Thus this light cone must be contracting
toward the future at $t=t_{i}$ and throughout the FRW region. Thus,
the part of any constant time slice (at $t>t_{i}$) visible to our
observer is metrically identical to a subset of that at $t=t_{i}$,
but with a lower entropy density (due to the expansion). Thus, if
our scenario satisfies the light-cone holographic bound at $t=t_{i}$, it will
do so throughout the entire spacetime. 

To identify the observer's past light cone at $t=t_{i}$, recall that
we assumed homogeneity, isotropy, and spatial flatness. Thus, the
metric takes the form\[
ds^{2}=dt^{2}-a(t)^{2}dx_{i}^{2}\]
and the Friedmann equation is 

\[
\left(\frac{\dot{a}}{a}\right)^{2}=\frac{\rho}{3}+\frac{\lambda}{3}\]
in units where the reduced Planck mass is set to one ($M_{pl}=(8\pi G)^{-1/2}=1$).
For matter satisfying $p=\kappa\rho$, the density satisfies (see
e.g. \cite{FLP})\begin{equation}
\rho\propto\frac{s_{i}^{1+\kappa}}{a^{3(1+\kappa)}},\label{eq:dendep}\end{equation}
 where $s_{i}$ is the entropy density in comoving coordinates, at
time $t=t_{i}$. The entropy density $s_{i}$ is constant during the
FRW phase. At the end of inflation $(t=t_{i})$ the scale factor is
$a_{i}=e^{N_{i}}\Lambda^{-1/2}$, where $N_{i}$ is the number of
$e$-foldings during inflation. Note also that inflation ends when
$\rho_{matter}$ is of order $\Lambda$. The universe then expands
in a Friedmann-Robertson-Walker phase dominated by matter with some
particular value of $\kappa$ until the residual cosmological constant
$\lambda$ comes to dominate at $t\approx t_{f}$. The scale factor
at this time is determined by requiring the matter density to decrease
to of order $\lambda$, so using (\ref{eq:dendep}) we find\begin{equation}
\frac{a_{f}}{a_{i}}\approx\left(\frac{\Lambda}{\lambda}\right)^{\frac{1}{3(1+\kappa)}}.\label{eq:arat}\end{equation}
Now consider the past light cone of a late-time observer. In the $\lambda$-region
the light-cone at each time encloses a spherical volume whose radius
$R_{f}$ at $t=t_{f}$ is of order the late-time de Sitter radius
$\lambda$. The requirement that each light ray is null ($ds^{2}=0$)
allows one to propagate the rays back in time and thus to determine
the size $R_{i}$ of the visible region at $t=t_{i}$. The result
is:\begin{equation}
R_{i}=\frac{a_{i}}{a_{f}}R_{f}+\frac{2}{3\kappa+1}\Lambda^{-1/2}\left(\left(\frac{a_{f}}{a_{i}}\right)^{\frac{3\kappa+1}{2}}-1\right).\label{eq:rinit}\end{equation}
Since $\Lambda\gg\lambda$ and any positive energy condition requires
$\kappa\ge-1$, (\ref{eq:arat}) and (\ref{eq:rinit}) yield\begin{equation}
R_{i}\approx\Lambda^{-1/2}\left(\frac{a_{f}}{a_{i}}\right)^{\frac{3\kappa+1}{2}}.\label{eq:riapprox}\end{equation}
The total entropy in this region is computed using (\ref{eq:dendep})
where, since $\rho=\Lambda$ at $t=t_{i}$, one finds\begin{equation}
S_{i}=R_{i}^{3}\Lambda^{\frac{1}{1+\kappa}}.\label{eq:entropyi}\end{equation}

Thus, the above version of the holographic bound is equivalent in
our context to the requirement \begin{equation}
S_{i}\lesssim\frac{1}{\lambda}.\label{eq:Sbound}\end{equation}
Inserting (\ref{eq:riapprox}) and (\ref{eq:entropyi}) yields\begin{equation}
\lambda^{-\frac{3\kappa+1}{2(1+\kappa)}}\lesssim\lambda^{-1},\label{eq:lambdaBound}\end{equation}
which is clearly independent of $N$. Now, since $\lambda<1$ in Planck
units, (\ref{eq:lambdaBound}) is equivalent to\[
\kappa\leq1,\]
and is the same bound that arises from causality considerations. Note
that although the region visible to the observer is restricted by
(\ref{eq:Sbound}), the entropy across the entire initial timeslice
$t=t_{i}$ can be arbitrarily large, allowing for an arbitrary number
of previous $e$-foldings. As described earlier, if the light-cone holographic
bound is satisfied at $t=t_{i}$, it will be satisfied at all times.
Thus no restriction of any sort arises from light-cone holographic considerations
in the spacetime of figure 1 and, in particular, light-cone holography does not
limit the number of $e$-foldings.

\section{Discussion}

We have established that light-cone holography places no bounds on the number
of $e$-foldings to the past of a late-time observer and is thus
consistent with the eternal inflation scenario.   This conclusion differs from that of
\cite{Banks:2003pt} because we do not share their assumption that local field theory can correctly compute the entropy of a volume larger than that contained in the past light-cone of any observer.  Again, we note
that assuming local field theory to correctly describe the entropy of
similar large volumes at late times would also contradict holographic
bounds.  In particular, the corresponding calculation in pure de
Sitter space would contradict the idea that asymptotically de Sitter
space has a finite number of states, on which the discussion of
\cite{Banks:2003pt} also rests\footnote{The authors of
  \cite{Banks:2003pt} express their skepticism of the existence of a
  consistent theory which approximates local field theory in the
  inflating $\Lambda$-region and leads to similar predictions for the
  CMB, but yields a smaller total entropy in the inflating regime.  We
  have no such example to offer, but see no reason why creating such a
  model is fundamentally more difficult than achieving the same goal
  for de Sitter space itself, a task not yet completed for which the
  authors of \cite{Banks:2003pt} expect success. See
  \cite{Guijosa,Verlinde} for some steps toward a model for de Sitter
  with a finite dimensional Hilbert space. Unfortunately, until a model exists for the perhaps simpler pure de Sitter case, there will be few solid grounds on which to resolve this difference of opinion.}.  Thus, at the current level of holographic understanding, we see no reason to suppose a contradiction between holography and a large number of $e$-foldings such as would arise in eternal inflation.

For eternal inflation to be self-reproducing, the inflaton must be able to fluctuate
up its potential with some finite probability, giving rise to inflating
regions with an increasing rate of expansion. One may also ask if
there are holographic constraints on this process.  Discussions of related issues have appeared in \cite{Coule}.

Let us begin with a clear example that illustrates how this mechanism
can be compatible with holography. Consider a region of spacetime
with some effective Hubble parameter $H$, homogeneous over many horizon
volumes. Suppose a bubble with effective Hubble parameter $H'>H$
is nucleated inside this region with a size larger than the horizon
size set by $H$. This process occurs with finite probability in the
eternal inflation scenario \cite{Linde:1994xx}. Applying the holographic
bound to this situation \cite{Easther:1999gk} one finds that the
generalized second law yields no constraint on the evolution of this
super-horizon size bubble, as it is unable to collapse. The bubble
is then free to expand in a manner compatible with holography, and
no contradiction is later reached when inflation ends in this bubble
and a vast amount of entropy is produced, despite the fact the bubble
started out near GUT scale size, with low entropy. From a quantum
mechanical viewpoint, the system starts in a special state of low
entropy, but as time evolves the state explores a larger subspace
of the full Hilbert space of states, corresponding to the $H'$ bubble
expanding into the ambient $H$ region. Clearly we have ignored inhomogeneities,
but we believe this example suffices to illustrate the essential compatibility
of the seeding mechanism of eternal inflation with holography.

Another oft discussed situation occurs when the initial radius of
the bubble $H'$ is smaller than the ambient spacetime's inverse Hubble
scale $H^{-1}$. For $H'>H$ this bubble will collapse and can form
a black hole whose interior becomes a baby universe that undergoes
inflation. For uncharged bubbles, Farhi and Guth \cite{Farhi:1987ty,Farhi:1990yr}
concluded the initial conditions for the formation of such a bubble
are always singular. This may prevent the formation of such bubbles
in the first place. Even if they are formed, a curvature singularity
separates any external observer from the inflating interior of the
bubble, so the application of holography is not entirely clear. The
case of charged bubbles was analyzed in \cite{Alberghi:1999kd}. There
it was found these problems can be avoided, but a new difficulty appears
because the inflating region lies inside a Cauchy horizon, which is
unstable \cite{SimpsonPenrose,ChandraHartle,PoissonIsrael,BurkoOri,Dafermos1,Dafermos2}. 

Let us nonetheless assume that such problems are somehow solvable
and that sub-horizon scale bubbles do play a role in seeding eternal
inflation. In this scenario, we wish to analyze possible constraints
of holography. If semi-classical physics is valid in the appropriate
regions of spacetime, the bubbles will collapse and form horizons.
Banks has argued the entropy of such bubbles should be bounded by
the Bekenstein-Hawking entropy of the resulting horizon \cite{BanksFalseVac}.
In particular, the argument is that universes such as ours today have
$S\approx10^{85}$, which requires an event horizon of radius $10^{8}m$
, somewhat larger than the radius of the earth. The probability of
nucleating such a large bubble in the early universe is extremely
small.

However there are a number of subtleties in the above argument. Let
us at least enumerate some of the possibilities, assuming we start
with a GUT scale bubble that collapses to form a black hole. Such
a GUT bubble might have formed during quantum fluctuations in the
eternal inflation foam, or perhaps through interactions in a very
high energy particle accelerator.

\begin{enumerate}
\item One interpretation of the Bekenstein-Hawking entropy $S_{BH}$ is
that $e^{S_{BH}}$ bounds the dimension of the Hilbert space of states
associated with the region inside the horizon. Let us denote this
Hilbert space by $\mathcal{H}$. This interpretation is supported
by string theory calculations of black hole entropy via D-brane methods.
At first glance, the idea that a black hole with initial Bekenstein-Hawking
entropy of order $S_{BH}\approx10^{6}$ expands to give a large universe
appears to conflict with this idea. In particular, suppose one assumes
that time evolution maintains a sharp distinction between those states
in $\mathcal{H}$ and those orthogonal to $\mathcal{H}$. By this
we mean both a) that $\mathcal{H}$ and its complement do not significantly
mix under time evolution and b) that that the two classes of states
appear quite different to local observers which experience them. In
this case, a local observer inside the bubble can estimate the dimension
of the Hilbert space of states similar to what she observes and compare
this with $\mathcal{H}$ for an initial GUT size bubble. While it
has been argued \cite{Banks:2002wr} that precise measurements are
impossible for this observer, it is clear that there are certain classes
of scattering observables that can be measured with very high precision.
The observer in our present universe would then be able to conclude
with a high degree of certainty that her Hilbert space is larger than
that inherited from a GUT scale bubble and rule out creation of her
universe via such a black hole .
\item However, it is not clear to what extent the assumption in (1) above
is physically justified. In particular, consider assumption (1a),
that the bubble remains in the space $\mathcal{H}$ under time evolution.
Certainly the original black hole Hawking radiates, and may well disappear
in the distant future. Black hole complementarity suggests the state
of the Hawking particles is actually equivalent to a state inside
the black hole, and in particular to any baby universe so created.
Since these Hawking particles explore a much larger Hilbert space,
it is conceivable then that the entropy of the bubble is not constrained
by holographic bounds. From the external point of view, the late time
de Sitter phase with large entropy could have a complementary description
as a high entropy state in the Hilbert space $\mathcal{H}\times\mathcal{O}$,
where $\mathcal{O}$ is the Hilbert space of states outside the horizon.
\item Let us also consider assumption (1b), that $\mathcal{H}$ and its
compliment appear quite different (for all time) to local observers
which experience them.  Such observers are unable to measure the exact
late-time state of the full system, and so end up measuring the entropy
of a locally accessible subsystem. It is not clear to us whether observations
of such subsystems can indeed distinguish between universes produced
via black holes and those which arose from other initial conditions.
Let us suppose now that they cannot. Let us also note that the von
Neumann ($Tr\rho\ln\rho$) entropy of such a subsystem may well \emph{exceed}
the entropy of the full system, because the observer is unable to
measure correlations with causally disconnected regions of the asymptotically
de Sitter space. Indeed the exact von Neumann entropy will vanish
if the system is in a pure state. It is thus conceivable that a late
time observer could see a vastly larger entropy than the Bekenstein-Hawking
entropy associated with the horizon of an initial black hole from
which our universe somehow emerged.
\end{enumerate}
We see that in order to arrive at a contradiction, one would need
to prove the existence of more than $e^{S_{BH}}$ states which a)
are macroscopically indistinguishable from our universe and b) could
have been formed from a GUT-scale black hole. We conclude that successful
production of a de Sitter region with large apparent entropy must
produce some fine-tuning of the universe, but not that it is otherwise
ruled out without additional assumptions. Such a fine tuning is not
a surprise, as the instability of the charged black hole's Cauchy
horizon and the resulting singularity already indicate that successful
production of a universe via a black hole is either far from generic
or is dependent on high energy effects not currently understood and
associated with the singularity. If one believes that the process
is possible at all, it is plausible that any fine-tunings required
by holography are a natural result.

In summary, we see that holography appears quite compatible with eternal
inflation. In particular, a late time observer sees no bound on the
number of $e$-foldings or on any other parameters in the model of
figure \ref{cap:Spacetime-diagram}. Furthermore, the mechanisms of
self-reproduction in eternal inflation survive holographic constraints.
Holography may place strong constraints on branches of the eternal
inflation spacetime that somehow emerge from black hole interiors,
but even here such a conclusion follows only if one introduces additional
assumptions. Because quantum gravitational processes are necessarily
involved in the production of such regions, any such assumptions are
necessarily difficult to test and must remain inherently speculative.

\begin{acknowledgments}
D.L. thanks T. Hertog and D. Lyth for a helpful discussions and KITP for hospitality.
D.L. is supported in part by DOE grant DE-FE0291ER40688-Task A and
NSF grant PHY99-07949. D.M was supported in part by NSF grants PHY00-98747,
PHY 99-0747, and PHY03-42749.
\end{acknowledgments}
\bibliographystyle{brownphys}
\clearpage\addcontentsline{toc}{chapter}{\bibname}\bibliography{eternal}

\end{document}